\documentclass[aps,article,groupedaddress]{revtex4}

\usepackage{graphicx}
\begin{document}
\bibliographystyle{apsrev}

\newcommand{\be}{\begin{eqnarray}}
\newcommand{\bb}[1]{
\textbf{#1}
}
\newcommand{\ee}{\end{eqnarray}}
\newcommand{\ds}[1]{
#1{\hskip-2.0mm}/
}
\newcommand{\lb}[1]{
\label{#1}
}
\renewcommand{\thefootnote}{\arabic{footnote}}
\setcounter{footnote}{0}
%---------------------------------------------------------------------------%

%\twocolumn[\hsize\textwidth\columnwidth\hsize\csname @twocolumnfalse\endcsname

\title{
{Parameter free calculation of hadronic masses
from instantons.}}
\author{Pietro Faccioli}
\address{Department of Physics and Astronomy\
State University of New York at Stony Brook/
Stony Brook, New York 11794, USA}
\date{\today}
\email{faccioli@tonic.physics.sunysb.edu}
%---------------------------------------------------------------------%

\begin{abstract}
We propose a non-perturbative calculation scheme 
which 	is based on the semi-classical
approximation of QCD and 
can be used to evaluate quantities of interest in hadronic physics. 
As a first application, we evaluate the mass of the pion and of the nucleon.
Such masses are related to a particular combination of Green functions
which,  in some limit,  is dominated by the contribution 
of \emph{very small-sized} instantons. 
The size distribution of these pseudo-particles
is determined by the 't Hooft tunneling amplitude formula 
and therefore our calculation is free from any model parameters.
We prove that instanton forces generate a light pion and a nucleon 
with realistic mass ( $M_n \sim 970\, MeV$ ).
In connection with sum-rules approaches, we discuss the overlap
of instantons with pion and nucleon resonances.

\end{abstract}
%\pacs{xxx}
\maketitle
%---------------------------------------------------------------------%
\section{Introduction}
\label{introduction}
After the discovery of the BPST instanton solution in QCD \cite{bpst},
two important results by 't Hooft 
originated a large interest in the semi-classical description of the QCD 
vacuum \cite{thooftzm1,thooftzm2}:
the calculation of the tunneling amplitude 
and the discovery of  
the existence of quark zero-modes in the field of an instanton (which allowed
to solve the U(1) problem)\cite{thooftu1}.

Unfortunately, it was immediately realized that in QCD (unlike in the 
electroweak case) the tunneling amplitude at one or two-loops accuracy is
 divergent in the infrared and therefore that
instantons ``tend to swell''.
This fact led some to the conclusion that the QCD vacuum should
be populated by large sized, topologically non trivial field configurations,
which cannot be described semi-classically.

On the other hand, in \cite{shuryak82} Shuryak argued that the divergence
 of the single
instanton density is non-physical and that the size distribution 
of instantons should eventually start to decay above some average size
 $\bar{\rho}$,
either due to non-perturbative effects (which are not included in the 
't Hooft formula) or due to interaction with other instantons \cite{dyakonov84}
(for a detailed 
discussion of this issues, see \cite{shuryakrev}).
Along this line, he concluded that a semi-classical approach can still 
justified if a small
fraction of the Euclidean space time is occupied by instantons,
i.e. if  $\bar{\rho}^4\,\bar{n}<<1$, where $\bar{n}$ is the average instanton
 density.
From an analysis of the global properties of the vacuum (quark and gluon 
condensate) he estimated these parameters to be
$\bar{n}\simeq 1\,fm^{-4}$, $\bar{\rho}\simeq 1/3\, fm$.
Based on such arguments,
the Instanton Liquid Model (ILM) was developed in the 80's and 90's.
The model was proven to be phenomenologically successful in describing many
important low energy properties of QCD, such as
the breaking of chiral symmetry \cite{dyakonovCSB},  
several hadronic masses and correlators \cite{shuryakrev} and, 
recently, electro-magnetic formfactors \cite{blotz,forkel,3ptILM,pionFF}.
Moreover,  studies of the classical content of the QCD vacuum, based on
lattice simulations via cooling procedure \cite{chu94}, confirmed the 
presence of topologically non 
trivial ``bumps'' of gauge fields, with typical density and size very close
to that assumed by the ILM.

Despite its phenomenological success, the ILM remains a model, 
because none has been able to derive from
QCD why large instantons are not present in the vacuum.
However, the fact that \emph{there are} small-sized instantons in the 
vacuum, whose distribution is described (at  two-loop level) by the
$\rho\,\Lambda_{QCD} <<1$ limit of the 't Hooft tunneling amplitude, is a
result  of QCD. 

In this work,   we try to set up  a calculation scheme for hadronic observables
 which emphasizes 
the contribution from such small-sized instantons, and we 
apply it to evaluate the mass of the pion and of the nucleon.
In practice, in order to keep our calculation finite at each intermediate 
stage,  we will need to regularize
the size distribution by  cutting-off large instantons.
However we shall show that, in an appropriate limit to be defined below, at the
end of the calculation our
results will not depend on the particular choice of the cut-off and the only
parameter left is $\Lambda_{QCD}$.

This way, we derive the existence 
of a light $(I=1,J^p=0^-)$ meson state (pion) and of a 
$(I=\frac{1}{2},J^p=\frac{1}{2}^+)$ baryon state (nucleon) .  
In the case of the nucleon, we are able to extract 
its mass and find $(3.9 \pm 0.2 )\,\Lambda_{PV}$. However, this framework
did not allow us to determine with sufficient accuracy the very small mass 
of the pion.

The paper is organized as follows. In section \ref{correlators} we introduce
our particular combination of two-point functions and show how this is related
to the masses of the corresponding hadrons.
In section \ref{calculation}, such a combination is evaluated
theoretically, in our semi-classical approach.
Results are presented and discussed in section \ref{discussion} and summarized
in section \ref{summary}.
 
%-----------------------------------------------------------------------%

\section{Determination of the hadronic masses from point-to-wall correlators
at large momentum}
\label{correlators}

Point-to-wall hadronic correlators are defined as the spatial Fourier 
transform of the point-to-point Euclidean correlation functions,
\be
\label{pw}
G^H_2(t,\bb{p})=\int d^3 \bb{x}\, e^{i\,\bb{p}\cdot\bb{x}}
\,<0|\,J_H(t,\bb{x})\,J^\dagger_H(0)|0>,
\ee
where $J_H$ is an interpolating operator, which excites states
 with the quantum numbers of the hadron $H$.
These correlators are often used to extract hadronic masses from lattice 
simulations.
In fact, by computing their spectral decomposition,
one finds that, in the large $t$ limit, the contribution
from all excitations is exponentially suppressed.
The energy of the ground  state can than be determined from the logarithmic 
derivative
\be
\label{logM}
-\frac{1}{dt}\,log\left[\frac{G^H_2(t+dt,\bb{p})}{G^H_2(t,\bb{p})}
\right]\stackrel{t\to\infty}{\rightarrow} E(\bb{p}).
\ee

By setting $\bb{p}=0$ in (\ref{logM}), one can directly extract the
mass of the lowest lying particle, with the quantum number of $H$.
In lattice simulations, $\bb{p}$ is typically chosen to be
 the smallest possible momentum compatible with the 
box size and  $dt$ to be the lattice spacing in the time 
direction.

In this work, however, we will need to  consider
ratio (\ref{logM}), for $|\bb{p}|\gtrsim 1\,GeV$.
In fact, we shall show that only in such kinematic range the pion
and nucleon correlators are dominated by the effects of
single  small-sized instantons, which are calculable theoretically.

As we increase the momentum, the determination of the masses
gets more and more difficult, and eventually becomes 
impossible if $|\bb{p}|>>m_H $.
This is due to the fact that, in the ultra-relativistic regime where 
$E(|\bb{p}|)\sim|\bb{p}|$,
all states in the spectral decomposition have the same exponential decay
\footnote{We acknowledge a useful discussion with A.Schwenk on this point.}.
As a consequence, as we increase the momentum, larger and larger  Euclidean
time intervals are needed, in order to isolate the ground-state.
Therefore, for a given value of the time interval $t$, the LHS of 
eq.(\ref{logM}) will be dominated by the ground-state only for a limited 
range of momenta.
Generally speaking, we expect the LHS of eq.(\ref{logM}) to become bigger than
the RHS, at some large momentum.  

The importance of the contribution form resonances depends on the details of
the spectral function.
In the case of the pion, we expect that they should become relevant only
 at very large momenta, because
the first resonance ($\pi(1330)$) is much heavier than the ground-state.  
To estimate such contribution, let us consider a very simple two-poles model, 
in which only $\pi$ and $\pi(1330)$ states are included.
Than the spectral decomposition of $G_2(\bb{p},t)$ would read:
\be
G_2(\bb{p},t)\simeq 
\frac{a_\pi}{2\sqrt{\bb{p}^2+m_\pi^2}}
e^{(-\sqrt{\bb{p}^2+m_\pi^2}\,t)} + 
\frac{a_{\pi(1330)}}
{2\sqrt{\bb{p}^2+m_{\pi(1330)}^2}}
e^{(-\sqrt{\bb{p}^2+m_{\pi(1330)}^2}\,t)},
\ee
where $a_\pi$ and $a_{\pi(1330)}$ are the coupling of the states 
with the interpolating operator (which we shall assume to be of
 the same order of magnitude).
For $t=7\,GeV^{-1}$ and $|\bb{p}|\simeq 2 \,GeV$ we have that the contribution
of $\pi(1330)$ is suppressed by one order of magnitude.

The nucleon channel is more problematic because the first resonance 
is only about $500 MeV$ heavier than the ground state.
In this case, by repeating the above analysis, we find that the desired
 suppression should be achieved for $t\sim 8\, GeV^{-1}$ and 
$|\bb{p}|\simeq 1.5\,GeV$ 
Above such thresholds, a theoretical analysis of (\ref{logM}) can 
be used to obtain qualitative information about the strength of 
the resonances and the continuum. 

%-------------------------------------------------------------------------%

\section{Theoretical calculation}
\label{calculation}

In this section, we evaluate the LHS of (\ref{logM}) 
in a semi-classical approximation.
Our starting point is the tunneling amplitude for one instanton, which
is derived by integrating over small quantum fluctuations about the instanton
 solution. At  two-loop accuracy it reads:
\be
\label{dens}
d_{'t Hooft}(\rho)=const\,\cdot\rho^{-5}\,\rho^{N_f}\,\beta_1(\rho)^{2\,N_c}\,
exp\left[-\beta_2(\rho)+\left(2\,N_c-\frac{b'}{2\,b}\right)\,
\frac{b'}{2\,b\,\beta_1(\rho)}\,ln[\beta_1(\rho)] \right],
\ee
where $\beta_1(\rho)$ and $\beta_2(\rho)$ are the one and two-loop beta
functions:
\be
\beta_1(\rho)=-b\,\ln(\rho\Lambda_{PV}),
\ee
\be
\beta_2(\rho)=\beta_1(\rho)+\frac{b'}{2\,b}\,
ln\left(\frac{2}{b}\beta_1(\rho)\right),
\ee
\be
b=\frac{11}{3}N_c-\frac{2}{3}N_f,\qquad
b'=\frac{34}{3}\,N_c^2-\frac{13}{3}N_c\,N_f+\frac{N_f}{N_c}.
\ee
This result was carried out in the Pauli-Villars regularization scheme and
 $\Lambda_{PV}$ is the corresponding scale parameter.

This formula is derived perturbatively and therefore 
is valid only for small instantons, for which $\rho\,\Lambda_{QCD}<<1$. Its 
infrared divergence indicates that 
a fully perturbative description of a single tunneling amplitude 
is inconsistent. This is the so called 'infrared catastrophe' of the instanton
calculus in QCD.

In the ILM, this problem is overcome by assuming that the vacuum can be 
described as an ensemble of interacting instantons, and that the single
instanton size distribution is cut from above by an average 
instanton-instanton repulsion.
The model depends on one  dimensionless
parameter, the instanton packing fraction $\bar{\rho}^4\,\bar{n}$, which can
not be derived and has to be estimated phenomenologically. 

Such an approach necessarily introduces some additional approximations.
First of all, in the presence of non negligible overlapping, 
the sum of many instantons is no longer an exact solution of 
the classical equation of motion.
Moreover,  the tunneling events 
can non be treated as completely independent and therefore it is not 
guaranteed  \emph{a priori} that the contributions of
such classical configurations
is not overwhelmed by quantum fluctuations.
However, one should mention that the ILM gives indeed a very good 
phenomenology,  therefore one can argue, \emph{a posteriori},   
that the assumptions made are legitimate.

Here, we try to go one step further and 
set up a calculation scheme
which is sensitive only to the small-size tail of the 
instanton density and contains no phenomenological parameters.
We shall only assume that the instanton packing 
fraction is small enough for the pseudo-particles not to loose 
completely their individuality. In other words, the QCD vacuum will still be
described as some instanton ensemble. 
However, we will not need to choose a particular value
for $\bar{\rho}$ and $\bar{n}$, nor we will have to model the 
instanton-instanton interaction.
As a matter of fact, we shall also not require that the semi-classical
approach is justified for \emph{all} pseudo-particle in the ensemble, because
our results will be sensitive only to the contribution from the 
small ones. 

The basic simplification introduced by the semi-classical approximation
is that the infinite dimensional integration over all gauge 
configurations is replaced
by a finite number of integrals over the collective coordinates of each 
pseudo-particle.         

Let us consider, for sake of definiteness
\footnote{The derivation can be straightforwardly generalized to 
all correlation
functions receiving contribution from \emph{at least two} 
zero-mode propagators.},
 the pseudo-scalar two-point 
correlation function, interpolated by the current:
\be
\label{ps} 
J_5(x)=\bar{u}(x)\,\gamma_5 d(x).
\ee
In our approach,  the two-point correlator is approximated
 with (for two flavors):

\be
\label{instcorr}
G(x,y):=<0|\,J_5(x)\,J^\dagger_5(0)|0>\simeq
\frac{1}{Z}\,
\int d \Omega f(\Omega)\,
\left[\prod_{f=u,d}\,det(\ds{D}_\Omega+m_{f})\right] 
Tr[ \gamma_5 S(x,0;\Omega) 
\gamma_5 S(0,x; \Omega) ],
\ee
\be
\label{Z}
Z=\int\, d  \Omega f( \Omega)\,\left[\prod_{f=u,d}\,det(\ds{D}_\Omega+m_{f})\right],
\ee
where $\Omega=(\Omega_1,...,\Omega_{N^+},\Omega_{N^+ +1},...,\Omega_{N_I+N_A})$ is the set of all collective
coordinate identifying the configurations of the instanton ensemble
\footnote{We shall assume that the first $N^+$ components of $\Omega$ will
span the moduli space of instantons, while the next $N^-$ components that of
anti-instantons.},
 $f(\Omega)$ is the weight
associated to a particular configuration $\Omega$ in the 
pure Yang-Mills theory \footnote{$f(\Omega)$ incorporates the effects of small
quantum fluctuations around the classical solution.} 
and $S(x,y;\Omega)$ is the quark propagator  in the configuration $\Omega$.
The quark propagator can be formally written as:
\be
S(x,y;\Omega)=\sum_\lambda \frac{\psi^\Omega_\lambda(x)
\psi^{\Omega\,\dagger}_\lambda(y)}
{-i\,\lambda + m},
\ee
where $\psi^\Omega_\lambda(x)$ are eigenvalues of the Dirac operator,
$\ds{D}^\Omega\psi^\Omega_\lambda=\lambda\,\psi^\Omega_\lambda$.
In the field of a single instanton, the propagator is known exactly  
\cite{brown78} and consists  of a non zero-mode part
 and a zero-mode part:
\be
S(x,y;\Omega_i)=S_{nzm}(x,y;\Omega_i)+S_{zm}(x,y;\Omega_i),
\ee
where $\Omega_i$ denotes the set of collective coordinate relative to
 the particular instanton $i$.
It can be verified that, when correlators receive contribution from more that
one zero-mode propagator ( like in the case we are considering),
the non zero-mode part can be very well approximated
with the free propagator:
\be
S(x,y;\Omega_i)=S_{0}(x,y)+S_{zm}(x,y;\Omega_i).
\ee

The propagator in the field of many instantons was evaluated in 
\cite{dyakonov86} and  consists of a 
non zero-mode part 
(which can again be approximated with a free propagator) and a
quasi zero-mode part, which can be expanded in the functional space
of zero-modes of individual instantons 
\footnote{The coefficients $C_{I,J}$ correspond
to the matrix elements of the inverse of the ``overlapping matrix'' 
(see \cite{dyakonov86} for details).}:
\be
\label{prop}
S_{zm}(x,y;\Omega)=\sum_{I,J} C^\Omega_{I\,J} \,\psi^{\Omega_I}_0(x) 
\, \psi^{\Omega_J\,\dagger}_0(y),
\ee
where $\psi^{\Omega_I}_0(x)$ are the well known zero-mode wave functions 
\cite{thooftzm1,thooftzm2}.

Inserting (\ref{prop}) in (\ref{instcorr}) we find:
\be
G_2(x,y)-G_{2\,free}(x,y)=
\frac{1}{Z}\left(
\sum_{I\,J=1}^{N^+ + N^-}\,
\sum_{I'\,J'=1}^{N^+ + N^-}
\int\,d\,\Omega \, f(\Omega)\,
\left[\prod_{f=u,d}\,det(\ds{D}_\Omega+m_{f})\right] 
C^\Omega_{I\,J}\,C^\Omega_{I'\,J'}\right.\cdot
\nonumber\\
\left.
\cdot Tr[\gamma_5\,\psi^{\Omega_I}_0(x)\,
\psi^{\Omega_{J}\,\dagger}_0(0)\,\gamma_5\,\psi^{\Omega_{I'}}_0(0)\,
\psi^{\Omega_{J'}\,\dagger}_0(x)]\right) \nonumber\\
\ee
where $N^{+(-)}$ denotes the total number of instantons (anti-instantons).
It is convenient to  re-write this sum in a such a way that the contribution 
of a single instanton (anti-instanton) is isolated:
\be
G_2(x,y)-G_{2\,free}(x,y)=\frac{1}{Z}\left(
N^+\int d\,\Omega f(\Omega)\,
\left[\prod_{f=u,d}\,det(\ds{D}_\Omega+m_{f})\right] 
|C^\Omega_{1\,1}|^2
Tr[\gamma_5\,\psi^{\Omega_{1}}_0(x)\,
\psi^{\Omega_{1}\,\dagger}_0(0)\,\gamma_5\,\psi^{\Omega_{1}}_0(0)\,
\psi^{\Omega_{1}\,\dagger}_0(x)]+\right.\nonumber\\
\left.
+ N^-\int d\,\Omega f(\Omega)
\left[\prod_{f=u,d}\,det(\ds{D}_\Omega+m_{f})\right] 
|C^\Omega_{N^++1\,N^++1}|^2
Tr[\gamma_5\,\psi^{\Omega_{N^++1}}_0(x)\,
\psi^{\Omega_{N^+ +1}\,\dagger}_0(0)\,\gamma_5\,\psi^{\Omega_{N^+ +1}}_0(0)\,
\psi^{\Omega_{N^+ +1}\,\dagger}_0(x)]\right)+...,\nonumber\\
\ee
where the ellipses denote terms which depend on the collective coordinates
of  more than one (anti)instanton.
Such terms will introduce effects which are suppressed
by powers of the instanton density and are typically sub-leading, as will
be discussed later.
Now we can integrate out the remaining $N^+ + N^- - 1$ components of $\Omega$
and find:
\be
\label{almost}
G_2(x,y)-G_{2\,free}(x,y)=\left(
\int d\,\Omega_1 \tilde{f}(\Omega_1)\,\frac{1}{m*^2}
Tr[\gamma_5\,\psi^{\Omega_{1}}_0(x)\,
\psi^{\Omega_{1}\,\dagger}_0(0)\,\gamma_5\,\psi^{\Omega_{1}}_0(0)\,
\psi^{\Omega_{1}\,\dagger}_0(x)]+\right.\nonumber\\
\left.
+ \,\int d\,\Omega_{N^++1} \tilde{f}(\Omega_{N^++1})\,\frac{1}{m*^2}
Tr[\gamma_5\,\psi^{\Omega_{N^++1}}_0(x)\,
\psi^{\Omega_{N^+ +1}\,\dagger}_0(0)\,\gamma_5\,\psi^{\Omega_{N^+ +1}}_0(0)\,
\psi^{\Omega_{N^+ +1}\,\dagger}_0(x)]\right)+...,\nonumber\\
\ee
where we have introduced the single-instanton distribution
\footnote{The definition of the single anti-instanton is, of course, 
completely equivalent.},
\be
\tilde{f}(\Omega_1)=N^+/Z\,\int d \Omega_{2}...d\Omega_{N^+}
f(\Omega)\,\left[\prod_{f=u,d}\,det(\ds{D}_\Omega+m_{f})\right]
\,|C_{1\,1}|^2\,m^{*\,2} 
\ee
and the mass parameter $m^*$ has been added to keep track of the 
dimensionality.

\begin{figure}[h|t,clip=]
\includegraphics[height=6.0cm,width=7.0cm,clip=]{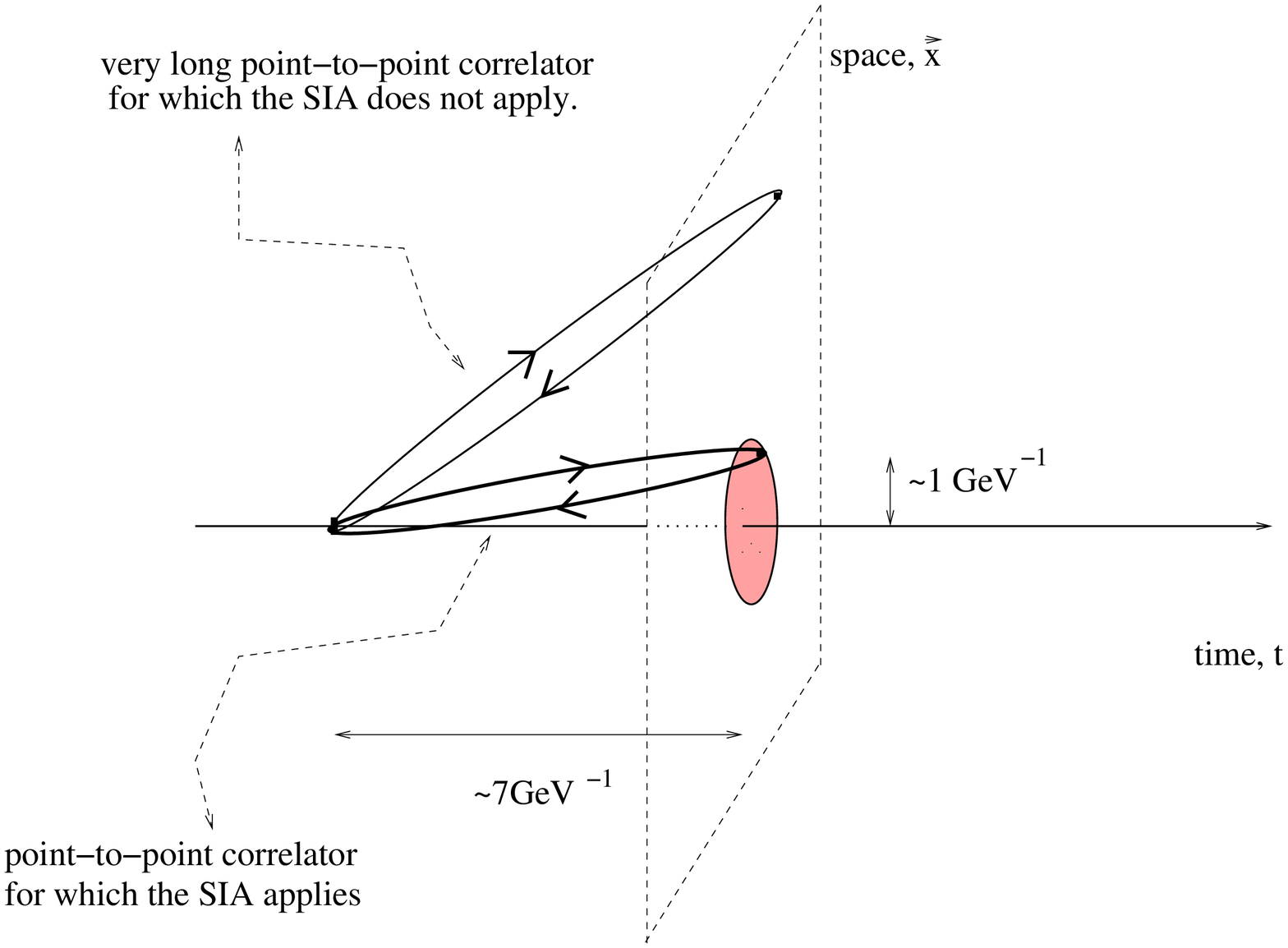}
\caption{Contribution of different point-to-point correlators. For 
$\bb{p}\sim 1\,GeV$ only the correlators with one end-point in the shaded area
contribute. For such correlators, the SIA is supposed to hold.}
\label{figSIA}
\end{figure}

Now let us now discuss the further simplifications that arise by applying just 
symmetry principles.
First of all, since the correlator we are considering is gauge invariant,
the integration over the color orientation of the instanton and of 
the anti-instanton is completely trivial.
Moreover, 
total momentum conservation
requires that the distribution of such instanton in the vacuum is homogeneous
\footnote{The implication of total momentum conservation in SIA will be 
discussed in detail in \cite{pionFF}.},
that is:
\be
\label{shuryakguess}
\tilde{f}(z;\rho)=\bar{n}\,d(\rho)
\ee
Notice that we have completely recovered the structure 
the single instanton density introduced by Shuryak in \cite{shuryak82}, 
who proposed the very simple ansatz:
\be
\tilde{f}(z;\rho)=\bar{n}\,\delta(\rho-\bar{\rho}),
\ee
which emphasizes the global properties of the instanton vacuum
(i.e. average instanton size and density).

In our calculation, however, we want to focus on the effect of small 
instantons.
At this purpose, we  choose a parameterization that, in the limit 
$\rho\to 0$, reproduces the tunneling amplitude (\ref{dens}).
For example we can choose the following regularization: 
\be
\label{nrho}
d(\rho)=d_{'t Hooft}(\rho)\,\frac{1}{1+exp((\rho-\bar{\rho})/\sigma)}.
\ee
We can now rewrite (\ref{almost}) as:
\be
G^{zm}_2(x,y)=G_2(x,y)-G_{2\,free}(x,y)=\bar{n}\left(
\int d^4 z\,\int_0^\infty d\,\rho \, d(\rho)\,Tr[\gamma_5\,S^I_{zm}(x,0;z,\rho)\gamma_5\,
S^I_{zm}(0;x;z,\rho)]+
\right.\nonumber\\
+ \left.\int d^4 z\,\int_0^\infty d\,\rho \, d(\rho)\,Tr[\gamma_5\,S^A_{zm}(x,0;z,\rho)\gamma_5\,
S^A_{zm}(0;x;z,\rho)]\right),
\ee
where we have taken equal density of instantons and anti-instantons, 
$n_I=n_A=:\bar{n}$, and  $S^{I(A)}_{zm}(x,y;z)$ 
is the zero-mode propagator of a particle of
 mass $m^*$ in the field of an instanton (anti-instanton) of size $\rho$ 
at $z$. 
Hence, we can identify the parameter $m^*$
as the quark effective mass originally introduced in 
\cite{shuryak82}. In \cite{sia} such parameter
was  estimated from numerical simulations
and form phenomenology to be $m^*\simeq 85 MeV$.

Summarizing, what we have done so far is to provide a formal derivation of
 the Single Instanton Approximation (SIA), starting from the partition 
function (\ref{Z}).
The validity of such approximation, in the case of the ILM,
 was studied in detail in \cite{sia}.
It that analysis, we considered several point-to-point correlators 
that receive contribution from two zero-mode propagators
and compared SIA calculations (with the ansatz (\ref{shuryakguess}))
 with numerical simulations in
the Random and Interacting Instanton Liquid.
We found complete agreement up to distances of the order of $6-7\, GeV^{-1}$,
depending on the particular correlator.
Since this value was obtained in a particular model, it should be considered 
as a rough  estimate.
We may argue that, if a particular observable is mainly 
sensitive to the contribution
of small instantons, the SIA should somehow increase its
range of validity, because the packing fraction 
of such pseudo-particles is smaller.
\begin{figure}[h|t,clip=]
\includegraphics[height=6.0cm,width=7.0cm,clip=]{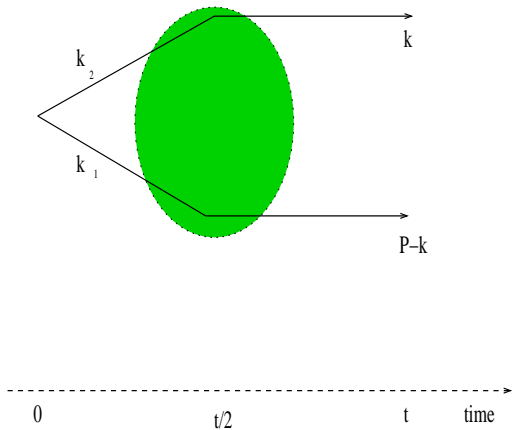}
\caption{Interpretation of the single small instanton domination of the 
point-to-wall correlators.}
\label{analogy}
\end{figure}
Also, we do not expect such SIA calculation for the point-to-wall propagator
 to be reliable for all values of the momentum $\bb{p}$.
In fact, if such momentum is small, the Fourier transform (\ref{pw}) 
receives non-negligible contribution
from  point-to-point correlators which connect 
the origin with points on the wall
that are very far away from the time axis (see fig. \ref{figSIA}).
From the discussion above it follows that such point-to-point 
correlators have size 
larger than the maximal compatible with the SIA.
However,  for $|\bb{p}|$ of the order of a $GeV$,  only points at the distance
smaller than roughly one inverse $GeV$ from the time axis 
will contribute to the Fourier transform, and the SIA is applicable.

Summarizing, the feasibility of our approach relies on the existence
of a window in the
time and momentum variables, such that the SIA is still holding and the 
ground-state is not overwhelmed by resonances.
For example, for $t\gtrsim 9\, GeV^{-1}$ and $p\lesssim 1 GeV$, 
we do not trust our approximation, while
for $t< 6\, GeV^{-1}$ and  $p\lesssim 1 GeV$, 
 the SIA would be reliable, but the correlation functions would start 
to become very sensitive to the contribution of the resonances. 
Based on such arguments, we shall choose $t$ to be
$7\, GeV^1 \lesssim t \lesssim \, 8\, GeV^{-1}$ and $|\bb{p}|\gtrsim 1\, GeV$.

Once the general calculation scheme has been established,
we can go ahead and evaluate the LHS of (\ref{logM}) for pion
and nucleon.
We begin with the pion pseudo-scalar point-to-wall correlation function.
The leading instanton contribution to $G_2^{\pi}(t,\bb{p})$ 
is given by:
\be
\label{pion}
G^{zm\,\pi}_2(t,\bb{p}) = \frac{\bar{n}}{4\,m^{*\,2}}\,
\int_0^\infty\,d\rho\,d(\rho)\,\rho^4 
\int_{-\infty}^\infty\,d\,z_4\,
e^{-|\bb{p}|\,\left[\sqrt{\xi(z_4-t)}+\sqrt{\xi(z_4)}\right]}\,
\frac{(1 +|\bb{p}|\,[\xi(z_4-t)]^{1/2})}{[\xi(z_4-t)]^{3/2}}
\,\frac{(1 +|\bb{p}|\,[\xi(z_4)]^{1/2})}{[\xi(z_4)]^{3/2}},
\ee 
where $\xi(x)= x^2+\rho^2 $.

The stationary phase method can also be applied to determine what 
instantons contribute the most, as  $|\bb{p}|$
increases.
By expanding the exponent around its minimum ($\rho=0$, $z_4= t/2$), it 
is straightforward to verify that, at large momenta,
 $G_2(t,|\bb{p}|)$ will be dominated by small-sized instantons.
Hence we expect that, in this regime, the prediction should be fairly 
insensitive to the detail of the cut-off function in (\ref{nrho}) and depend
essentially on $n_{'t Hooft}$, only.

The result that a single, small instanton dominates the point-to-wall 
correlation function at large momentum has an 
intuitive interpretation 
\footnote{We would like to thank E.V. Shuryak for suggesting such
interpretation.}. 
The current at origin emits a quark and an anti-quark with 
undetermined momenta (see fig. \ref{analogy}).
In order for such particles to form a bound-state at some later time $t$, 
their momenta has to be focused and made almost collinear.
The nearby instanton works as a ``lens'' and provides the required 
bending of the ``quark beams''.
At high momentum, the focusing is optimal if the ``curvature'' of the 
instanton lens is small
and sits at a distance $z_4= 1/2\, t$ from the source.

The same calculation can be repeated for the nucleon
point-to-wall correlation function.
In such case, we have more freedom in the choice of the point-to-point 
correlator.
Previous studies \cite{shuryakrev} have shown that the single-instanton 
contribution is largest in the correlator: 
\be
G^n_2(x,y)=<0|Tr[j_{sc}(x)\bar{j}_{sc}(0)\,\gamma_4|0>,
\ee
where $j_{sc}(x)=[u_a(x)\,(C\,\gamma_5)\,d_b(x)]\,u_c(x)\,\epsilon_{a\,b\,c}$
is the so called 'scalar current'.
The leading instanton contribution to the nucleon
 point-to-wall correlator reads:

\be
\label{nucleon}
G^{zm\,n}_2(t,\bb{p}) = \frac{192\bar{n}}{|\bb{p}| m^{*\,2} \pi^2}
\int_0^\infty dr\frac{r\,sin(|\bb{p}|\,r)}{[r^2+t^2]^{3/2}} 
\int_0^\infty d\rho\, d(\rho) \rho^4
\int_{-\infty}^\infty d z_4
\int_0^\infty d|\bb{z}| \frac{|\bb{z}|^2(r^2+|\bb{z}|^2+\xi(z_4-t))}
{[|\bb{z}|^2+\xi(z_4)]^2 [(r^2+|\bb{z}|^2+\xi(z_4-t))^2-4
|\bb{z}|^4\,r^4]^2}\nonumber\\
\ee

Again, we expect this result to be reliable only for $|\bb{p}|\gtrsim 1\,GeV$
 and that small instantons should dominate, in the large momentum regime.

One can check that, at the scale we are working at, the
free contribution to the pion and nucleon point-to-wall correlators is 
much smaller than the corresponding
leading instanton term,  for any reasonable value of $m^*$. 
Hence, in the ratio (\ref{logM}), the free contribution can
be safely neglected and the model parameters
$\bar{n}$ and $m^*$ cancel out. 
As a result, the only parameters that remain are
$\Lambda_{QCD}$, appearing in $d_{'t Hooft}(\rho)$ and the 
position of the instanton size cut-off $\bar{\rho}$.
However, from the above discussion,  we expect the results to become
independent on $\bar{\rho}$ for sufficiently large values of $\bb{p}$.

In this section we have focused on the evaluation of the logarithmic
derivative (\ref{logM}), which allows to extract hadronic masses.
In principle, the same calculation scheme can be applied to other
hadronic observables such as, e.g., electro-magnetic form factors 
\cite{pionFF}.
Following this prescription, one needs to relate the  
observable being
investigated to some ratio of point-to-wall correlation functions
and choose $t$ and $\bb{p}$ according to the above discussion.
One should keep in mind, however, that the SIA approximation 
is much more accurate if the relevant Green functions
receive contribution from more than one quark zero-mode propagator \cite{sia}.

%-------------------------------------------------------------------------%
\section{Results and discussion}
\label{discussion}
\begin{figure}[h|t,clip=]
\includegraphics[height=6.0cm,width=7.0cm,clip=]{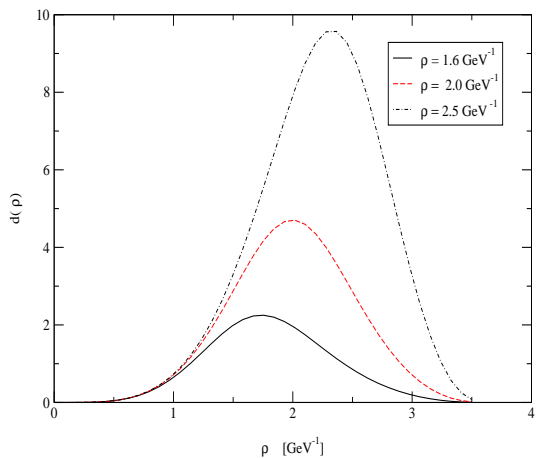}
\caption{The (unnormalized) single instanton tunneling amplitude for 
different values
of the cut-off $\bar{\rho}$ ($\Lambda_{PV}=250\,MeV$).}
\label{nrhofig}
\end{figure}
\begin{figure}[h|t,clip=]
\includegraphics[height=6.0cm,width=7.0cm,clip=]{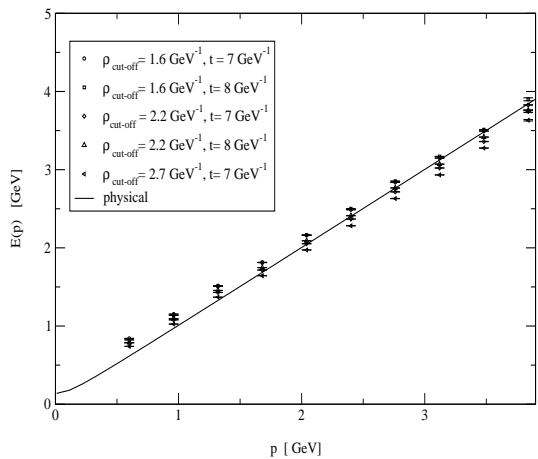}
\caption{Dispersion curve for the pion ($\Lambda_{PV}=250\, MeV$).
The dotted line is obtained from the physical pion mass, 
the points are SIA results.}
\label{pionresult}
\end{figure}

Let us begin by discussing the results obtained in the case of the pion
\footnote{In performing the calculation, we have expressed all dimensional 
quantities
in terms of $\Lambda_{PV}$.The results presented in this section have been 
converted to natural units by choosing $\Lambda_{PV}=250 MeV$.}.
In fig. \ref{pionresult},  several theoretical evaluations
 of the pion dispersion curve $E(|\bb{p}|)$, 
are compared with the physical one.
The different SIA  predictions were obtained from
(\ref{logM})  using 
different values of the time interval and of the size cut-off 
\footnote{These results were obtained using $\sigma= 0.1 GeV^{-1}$ 
in (\ref{nrho}). Different choices were tried and gave equivalent 
results.} 
  see fig. \ref{nrhofig}).
Notice that  these values correspond to extremely different pictures of the
 vacuum, because the corresponding
packing  fraction is proportional to $\bar{\rho}^4$ and 
changes by one order of magnitude. 

First of all we can verify that the pion pole has been isolated and that
the results are fairly
independent on the value of cut-off used.
From the comparison in the region $1\,GeV\lesssim |\bb{p}|\lesssim 2\,GeV$, 
we conclude that our semi-classical approach provides with the right amount of
dynamics and is completely consistent with the existence of a light pion.

Following the discussion of  section \ref{correlators}, we would 
expect the excited states to contribute to the relevant 
green-functions, in the region 
above $2\,GeV$. This implies that the logarithmic derivative (\ref{logM}) must
deviate from the pion dispersion curve, at some high momentum.
However, we observe that the complete agreement between our calculation
 and the pion dispersion curve continues up to  high momenta.
This fact can be interpreted in terms of quark-hadron duality
\cite{shifman80}, 
which assumes that the contribution from the continuum is
dominated by perturbative effects, while the ground-state pole
can be described in terms of non-perturbative effects.
Indeed our purely non-perturbative calculation does not reproduce
the presence of a continuum.
 
\begin{figure}[h|t,clip=]
\includegraphics[height=6.0cm,width=7.0cm,clip=]{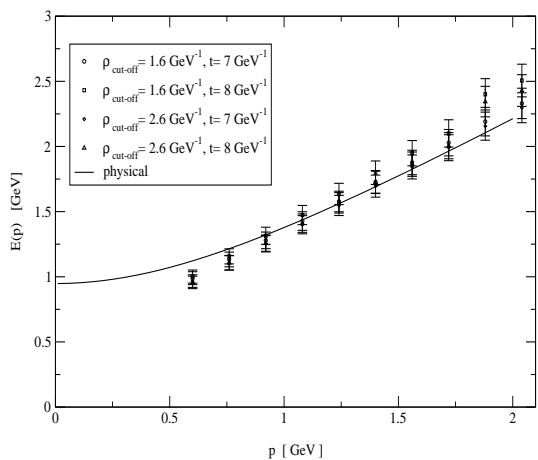}
\caption{Dispersion curve of the nucleon ($\Lambda_{PV}=250\, MeV$).
The solid line is obtained from the physical nucleon mass,
the points are SIA results. 
}
\label{nucleonresult}
\end{figure}

The results obtained in the case of the nucleon,
using different Euclidean time intervals and instanton size cut-offs, 
are reported in fig.  \ref{nucleonresult}.
The numerical solution of the integral (\ref{nucleon}) is quite 
challenging and the signal becomes very noisy above $2\,  GeV$.
Again, we verify that the ground state pole has been isolated and that the 
results are insensitive to the choice of the cut-off.
We observe agreement between our theoretical calculation and
the experimental curve in the expected region, $1\, GeV \lesssim |\bb{p}|
\lesssim 1.6\,GeV^2$ . 
Moreover, unlike in the case of the pion, the mass of the nucleon 
is sufficiently large do be extracted from the best fit curve
\footnote{According to the discussion in (\ref{correlators}), we have fitted
only the points in the region 
$1\, GeV <|\bb{p}|< 1.6\, GeV$.}.
We found $M_n=3.9 \pm 0.2 \Lambda_{PV}\sim 970 MeV$, 
consistent with the experimental value.

In this case, by extrapolating the trend of the calculations  above
$2\,GeV$, we may say that some little deviation from the
 nucleon dispersion curve seems to occur, at high momenta.
This fact may suggest that, unlike in the case of the pion, instanton 
non-perturbative effects have some overlap with nucleon resonances.

%-------------------------------------------------------------------%
\section{Conclusions and outlook}
\label{summary}
In this paper we have presented a new calculation scheme which can be
used to study several hadronic observables.
The method is based on the idea that, in the appropriate limit, it is possible 
to isolate the contribution form small-sized instantons, upon which we have 
theoretical control.

The prescription was applied to evaluate the mass of the pion and 
of the proton.
We found agreement with phenomenology and verify that
 our  results are insensitive to the particular choice
of  cut-off $\bar{\rho}$ and therefore are model independent.

By comparing the results obtained for the pion and nucleon, we
discussed  the importance of 
instanton effects for the physics of resonances, in these channels. 
We found that, in the case of the pion, a purely non-perturbative instanton
calculation does not reproduce the existence any resonance or of a continuum up
to extremely high momenta.
In the case of the nucleon, some deviation from the ground-state
dispersion curve seems to occur at high momenta (although
our results are not accurate enough to make a definitive statement).
This would imply
that instanton effects could have some overlap with nucleon resonances.
We intend to develop a more quantitative analysis of this issue in a future
work.

We think that the method could be useful in investigating the
contribution of instantons to a variety of hadronic observables.
An application of the same ideas to the analysis of pion and proton
 electro-magnetic  formfactors is in preparation.

%-------------------------------------------------------------------%
\begin{acknowledgments}
I am deeply indebted to E.V. Shuryak. 
I also would like to thank A. Schwenk for many useful discussions.
This work is partially supported by the US DOE grant No. DE-FG02-88ER40388.
\end{acknowledgments}
%-----------------------------------------------------------------------%

%\bibliography{instanton}

\end{document}